# Hardware Engines for Bus Encryption: a Survey of Existing Techniques

R. Elbaz, L. Torres, G. Sassatelli
LIRMM, UMR University of
Montpellier 2-CNRS C5506
Email: {`name@lirmm.fr`}

P. Guillemin, C. Anguille,
M. Bardouillet
STMicroeletronics
Advanced System Technology
Email: pierre.guillemin@st.com

C. Buatois, J. B. Rigaud
CMP, Centre microélectronique
de Provence Georges Charpak
Email: {`name@emse.fr`}

**Abstract**

*The widening spectrum of applications and services provided by portable and embedded devices bring a new dimension of concerns in security. Most of those embedded systems (pay-TV, PDAs, mobile phones, etc…) make use of external memory. As a result, the main problem is that data and instructions are constantly exchanged between memory (RAM) and CPU in clear form on the bus. This memory may contain confidential data like commercial software or private contents, which either the end-user or the content provider is willing to protect. The goal of this paper is to clearly describe the problem of processor-memory bus communications in this regard and the existing techniques applied to secure the communication channel through encryption – Performance overheads implied by those solutions will be extensively discussed in this paper.*

## 1. Introduction:

The range of services provided by every single embedded system tends to widen rapidly: banking transactions, web browsing, application / game download are nowadays common applications on mobile devices. In many cases, confidentiality in multiple forms has to be guaranteed: end-user private data must be kept secret, illegal software use must be avoided, etc. Unfortunately, the amount of memory needed for those platforms imply to use external memories for storing software and data, usually in clear form. That makes the processor-memory bus the weakest point of the system, hacker's favorite security hole. Observing both memory content and system execution can be done through simple board-level probing at almost no cost.

The obvious basic tool to counter these attacks is cryptography.

Cryptography can be divided in two families:
- Symmetric cryptography (a.k.a. private-key cryptography) [1]; the same key is used to cipher and decipher a message. Symmetric cryptography can be divided in two subfamilies, stream cipher (RC4, Steal) and block cipher (DES, 3DES, AES), both depicted later in this paper.
- Asymmetric cryptography (a.k.a. public-key cryptography) [1]; Two keys are used, one is public (known by every one) and the other is private and therefore has to be kept secret. This family of algorithms can address different purposes: if the private key is used to encrypt a message, everyone can decrypt it using the public key: such a scheme is usually used for authentication. On the other hand, if the public key is used to encrypt, only the private key owner can decrypt it, in this scheme, asymmetric cryptography is used for ciphering. That enables to achieve secure data exchanges over networks without having previously agreed of a common private key.

All cryptographic schemes are confronted to the temporal problem: the key must be long enough to thwart the "Brute force attack". These attacks consist in trying all possible keys. It's usually considered that a cryptosystem has a lifetime of at most 10 years due to the increase in computer processing power (Moore's law).

Two major issues are to be considered when dealing about security: *integrity* ensures that the data has not been modified (usually done for breaking into the system) while *confidentiality* guarantees the privacy of the data. This paper will only address the latter through existing encryption techniques. Moreover it will not explore the key management mechanisms relative to multitasking operating systems; refer to [2] for extensive discussion on that topic.

The paper is organized as follows:
Section 2 presents the context. Emphasis will be placed on the problem itself, and the consequences of inserting cryptosystems into SoCs will be discussed. Classification and targeted security level will also be exposed. Section 3 presents the different hardware encryption units proposed in the literature. Section 4 relates some yet unexplored solutions, discussing their respective advantages and drawbacks. Finally, some conclusions are drawn in section 5.

## 2. Context:

### 2.1 Software confidentiality

The emergence of on-demand software downloading services brings in the foreground the problem of content protection. It is therefore mandatory to establish a secure channel over a non secure network. As depicted in Figure 1 (issue from [3]) the software editor protects his product against piracy according to the following technique: a session key (K) is chosen and the software is ciphered using a symmetric algorithm. However in this case there are two potential risks:

i) A third-party may intercept the session key; the session key is the weak point and shall not be transmitted in clear form over the network (necessary prior to establish the secure channel). That's where public key algorithms come into play: they allow to transmit the session key in encrypted form; the protocol used is as follows:
1. A couple of keys – private ($D_m$) / public ($E_m$) – is provided by the chip manufacturer. The first one ($D_m$) is stored in a non-volatile memory inside the "secure" processor, and the second one ($E_m$) is transmitted to everyone who wants to communicate in a secure way with the processor.
2. The processor requests from the software editor the session key (K).
3. The software editor requests from the chip manufacturer the public key ($E_m$), who sends it over a non secure transmission channel.
4. The software editor sends, on the non secure transmission channel, a ciphered version of K, obtained with an asymmetric algorithm and ($E_m$).
5. Only the processor can decipher the message (K ciphered version) with the private key ($D_m$) stored in the non-volatile memory.
6. Finally, the processor uses K and a symmetric algorithm to decipher the software and to install the code in the external memory.

ii) The second risk is that the software might be copied afterwards: it is stored in clear form inside the external memory. During execution, information (data and instructions) are exchanged with the processor in clear: the bus can be probed and the software can be distributed without the knowledge of the software editor. The problem is shifted from the transmission channel to the communication bus between the CPU and external memory.

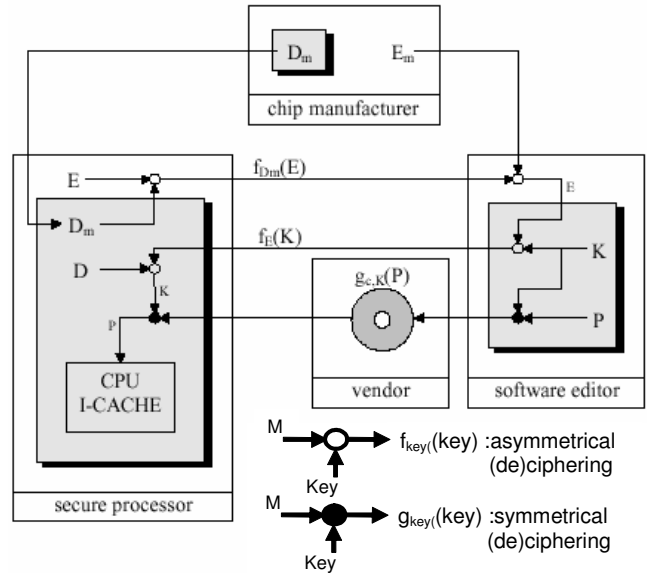

**Figure 1: Secret key exchange protocol on a non secure transmission channel**

In order to prevent that, the first idea would consist in using the symmetric algorithm employed by the software editor between the CPU and the external memory. However, the software editor can choose among different symmetric algorithm witch additionally might not be appropriate for processor-memory communication. Therefore, the deciphering step (with the session key (K)) often takes place in software.

The cipher unit has to be on the System On Chip (SOC), to complicate observation of the cryptosystem process, and thus, taking into account constraints such as: area, power consumption, performance penalties, is mandatory.

### 2.2 Cryptosystem SOC context

Electing a cryptosystem has to be done with respects to the system specifications. It is often a tradeoff between intended security (robustness) and affordable performance loss. The deciphering process implied by external CPU request will be deeply explored because of the usually stated critical impact on performance. Despite memory content ciphering can be done offline, data encryption has nevertheless to be considered as well for memory write operations, that will be discussed later in this paper.

*Asymetric vs Symetric cryptography*:
Asymmetric cryptography algorithms are often based on modular arithmetic, and operate on huge integers (512-2048 bits). They require more processing power (due to modular exponentiation) than symmetric algorithm for an equivalent robustness. Moreover, ciphered text is longer than the original clear text; larger memories are thus needed. Therefore, only symmetric algorithms will be considered in this paper.

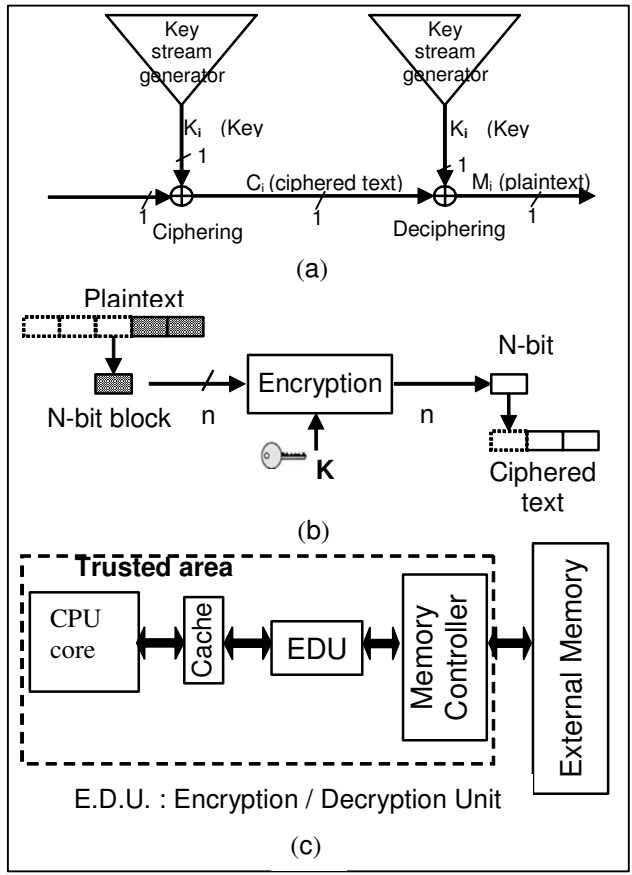

**Figure 2: (a) Stream cipher principle; (b) Bloc cipher principle; (c) Example of encryption unit placement**

Symmetric cryptography algorithms are divided in two families: stream cipher and block cipher.

Concerning block cipher, the plaintext is splitted in blocks (Figure 2b), and then each block is ciphered. On the other hand, the stream cipher principle (Figure 2a) is based on a XOR gate and a key stream. The first step is to generate the key stream. Then, the cipher text results from a XOR operation between the plaintext and the key stream. The robustness of this cryptosystem is based mainly on the key stream generation.

In our context, stream cipher seems to be more suitable in term of performance: the key stream generation can be parallelised with external data fetch. The shortcoming of block cipher cryptosystems is that deciphering cannot start until a complete block has been received.

The following observations must be taken into account by the cryptosystem designer; they're mainly related to block cipher implementations issues but can also apply to stream cipher to some extend.

Block cipher algorithms (symmetric algorithms) allow different ciphering mode [1]. Electronic CodeBook (ECB) is the most obvious mode; ciphered blocks is a function of the corresponding plaintext block, the algorithm and the secret key. Consequently a same data will be ciphered to the same value; which is the main security weakness of that mode. Cipher block chaining (CBC) mode provides improved security since each encrypted block depends also on the previous plaintext block. Its use proves limited in a processor-memory system due to the random data access problem (JUMP instructions).

If the encryption unit is located between cache memory and memory controller (Figure 2c), data stored in the cache memory will be in clear form. Accordingly, each cache miss implies external memory access along with additional latency due to the deciphering process. However, a write operation can have an even worst impact on the performance. The writing operation of a data smaller than the ciphered block size is penalizing because implies the following steps:
- Read the block from memory,
- Decipher it,
- Modify the corresponding sequence into the block,
- Re-cipher it,
- Write it back in memory.

One of the challenges of the cryptosystem design would be to hide these latencies.

### 2.3 Attack classification

IBM proposed a taxonomy [4] of adversaries and attacks in order to classify the security level achieved by each of their product: "*Adversaries were grouped into three classes, in ascending order, depending on their expected abilities and attack strengths:*
*Class I (clever outsiders): They are often very intelligent but may have insufficient knowledge of the system. They may have access to only moderately sophisticated equipment. They often try to take advantage of an existing weakness in the system, rather than try to create one.*
*Class II (knowledgeable insiders): They have substantial specialized technical education and experience. They have varying degrees of understanding of parts of the system but potential access to most of it. They often have access to highly sophisticated tools and instruments for analysis.*
*Class III (funded organizations): they are able to assemble teams of specialists with related and complementary skills backed by great funding resources. They are capable of in-depth analysis of the system, designing sophisticated attacks, and using the most sophisticated analysis tools* (very expensive). *They may use Class II adversaries as part of the attacks team.*"

Throughout this paper, the consumer market is targeted. For a hacker the cost of the attack should not exceed the price of the protected entity or the amount of profits expected. Thus, in our case, only attacks and adversaries classified in class II are taken into account.

Only none strongly invasive attacks are considered, physical access to data is limited to bus probing. The main objective is to prevent an attacker from understanding the contents of the data stored in external memory. Reference [5] provides a complete overview of existing attacks targeting secure embedded systems.

For example a famous attack carried out against the bus-encryption system of the DS5002FP microcontroller [6] by Marcus Kuhn consisted in the following:

The security principle of this microcontroller is based on a ciphering by block of 8-bit instructions. The hacker circumvents the cryptographic problem by finding a hole in the architecture processing and by applying exhaustive attack (8-bit instruction ⇔ 256 possibilities). After having identified the MOV instruction, he dumped the external memory content in clear form through the parallel-port.

## 3. State of the art:

The principle of bus encryption (encryption of the external memory content) was first introduced by Best 25 years ago [7] [8] [9]. Best proposed to consider the CPU as secure and consequently all data and addresses are in decrypted form inside the CPU and encrypted outside the SOC. Accordingly, a cipher unit is implemented on-chip, and a secret cipher key is located in an on-chip register (Figure 3). The block cipher chosen is based on basic cryptographic functions such as mono and poly-alphabetic substitutions and byte transpositions.

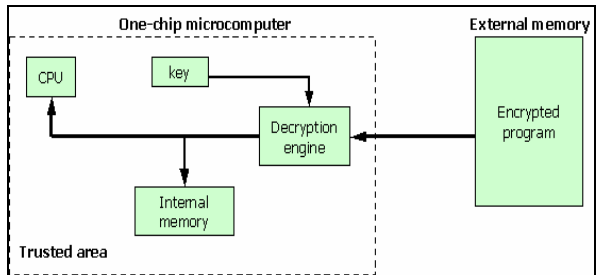

**Figure 3: Principle of Best's patent**

Some of the rules enounced by Best are still today considered as reference: System On Chip (SOC) is trusted, cipher unit and secret key remain on-chip, moreover all proposed hardware encryption units are located between the cache and the external memory controller.

Concerning industrial research, numerous patents exist. VLSI technology [10] proposes an architecture (Figure 4) where data transfers to and from the external memory are done page-by-page. All CPU external requests are managed by a secure DMA unit and communications between external and internal memory use an encryption / decryption core. This system allows the use of block cipher techniques (robustness). As the DMA is controlled by the operating system, this technique is viable provided that the OS is trusted.

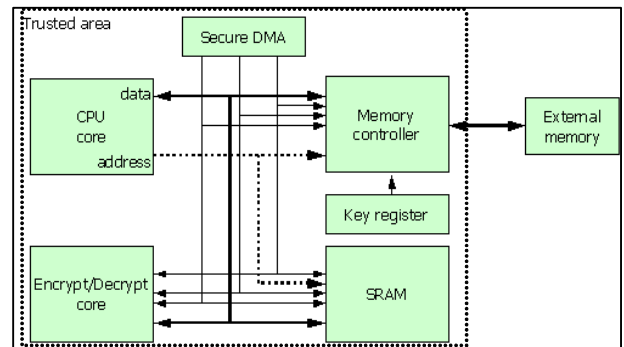

**Figure 4: Principle of VLSI technology's patent**

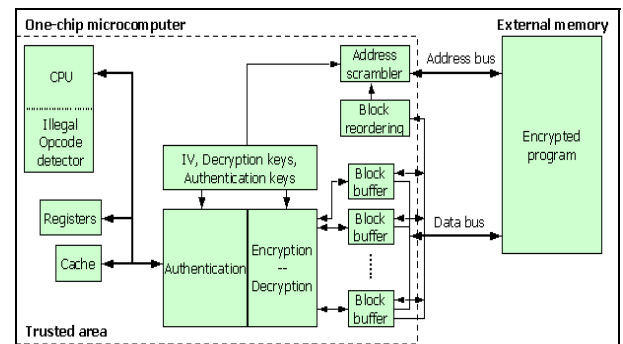

**Figure 5: Principle of the General Instrument's patent**

Another patent, by General Instrument Corporation [11], proposed to encrypt the memory content with a 3-DES (Data Encryption Standard) (Figure 5) in block chaining mode (CBC), and to offer the possibility to authenticate the data coming from external memory thanks to a keyed hash algorithm. Nonetheless, as seen previously, cipher block chaining technique is very robust but implies unacceptable CPU performance degradation for random accesses in external memory.

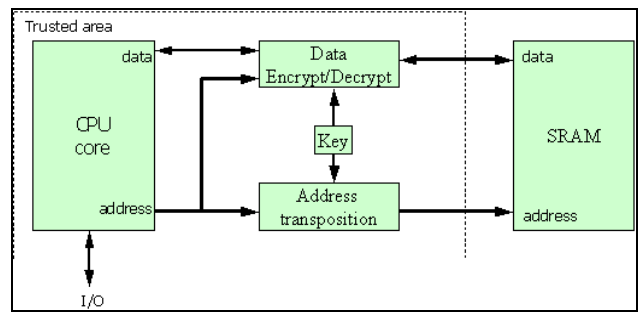

**Figure 6: Principle of the Dallas semiconductor's device**

A famous industrial device designed for different markets like pay-TV access control and credit cards is the one proposed by Dallas Semiconductor (Figure 6). The old version of their product, the DS5002FP [12], was broken by the well known Markus G. Kuhn attack [6]. The new one, the DS5240 [12], implements a ciphering based on a true DES or 3-DES block cipher which strengthened the robustness of the product. Accordingly, the 8-bit based ciphering passes to 64-bit based ciphering.

Concerning academic works, three projects [3][13][14] proposed tamper-resistant architectures including bus encryption units.

Guilmont et al. [3] use a fetch prediction unit and pipelined triple-DES block cipher. They assume to keep the deciphering cost under 2,5% in term of performance cost. However, this work only addresses static code ciphering and consequently authors are not confronted to smaller-than-block-size memory operations (occurs for data writing). As seen previously, great performance loss is to be expected in such cases.

The Xom project [13] uses a pipelined AES (Advanced Encryption Standard) block cipher as cipher unit which features a low latency of 14 latency cycles, while a throughput of one encrypted/decrypted data per clock cycle is claimed. In these projects the cipher unit is not the central focus: it's a part of a tamper-resistant architecture, and no specific benchmark has been carried out. Indeed, taking into account only the latency doesn't inform about the overall system cost.

The third work is called AEGIS [14]. In this work, the bus encryption engine is evaluated separately, therefore the given cost overhead information in terms of performance and silicon area are more detailed. The cipher unit is composed by a pipelined AES (300,000 gates) in CBC (Cipher Block Chaining) mode. But the ciphering block chain corresponds to a cache block, thus allowing random access to external memory (each cache block may be ciphered in CBC mode separately). However, the fetch instruction cannot be provided to the processor until an entire cache block is deciphered. The generation of the initialization vector (IV) needed by the CBC mode proves really secure: it is composed by the block address and by a random vector; to thwart the birthday attack [1] it is possible to replace the random vector by a counter. The major drawback is again the heavy trade-off between security and embedded constraints: they estimate the performance overhead induced by the encryption engine to 25%.

Numerous researches were carried out on the subject; the principle allowing a strong security is known: hardware implementation of algorithm approved by the NIST [15] – National Institute of Standard and Technology – in the past DES algorithm and currently the AES; however the cost overhead generated by those implementations remains considerable and unacceptable (in particular in term of execution time).

## 4. Various solutions considered:

All the presented existing solutions proposed a cipher unit on chip between cache memory and controller memory (Figure 7a).

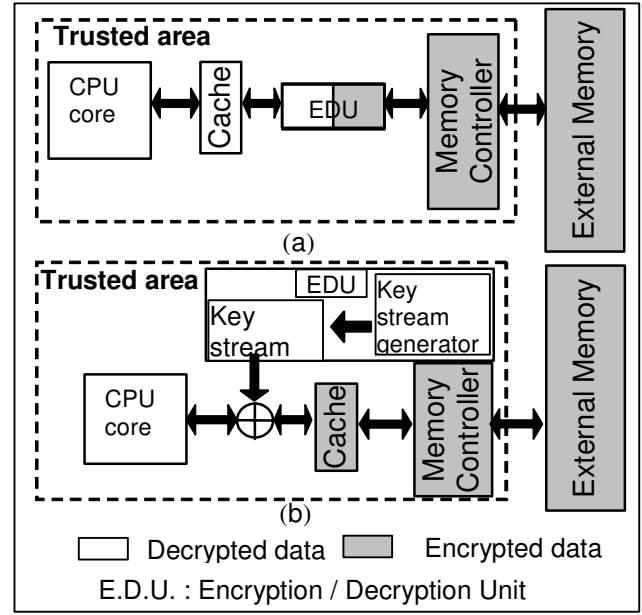

**Figure 7: (a) EDU between cache and memory controller; (b) EDU between CPU and cache memory**

Another possibility is to use a cipher unit during the data transfers between CPU and cache memory (Figure 7b). Consequently, all the data contained in the cache memory will be ciphered.

However, this scheme is critical. Modifying the cache access time directly impacts the system performance. The encryption / decryption process must be transparent for the CPU. That's why the cipher unit shall be a stream cipher. Additional problems appear: the key stream must be available on-chip to prevent drastic performance loss; it must also be sufficiently random to be secure.

Concerning the availability of the key stream and its reproduction for the deciphering process, a simple solution to resolve these two problems is to store it in an on-chip memory. That implies to add an on-chip memory equivalent to the cache memory in term of size.

Ciphering cache by this way, seems to enhance the security level (against the class III attackers which can observe an integrated memory); however, if the monitoring of on-chip memory is considered feasible, the hacker can easily obtain the key stream (stored in an on-chip memory) and can easily decipher the cache content without difficulty; consequently a solution must be found to protect the key stream if the security level has to fend off class III attackers. Concerning the key stream generation, several schemes are possible, but one constraint has to be matched: the time to create the key stream corresponding to a cache line must be equal, in the worst case, to an external memory data fetch otherwise it again implies important performance loss. Indeed, the generation of the key stream corresponds to a cache miss after a CPU data request.

Moreover, this scheme seems to provide no benefit in term of performance when compared to a stream cipher located between cache memory and memory controller.

The problem of an encryption unit is mainly the CPU performance degradation. A possible solution to improve performance would be to add a compression step to a ciphering solution (Figure 8). The compression has to be done before ciphering, if not, compression will have a very poor ratio due to the strong stochastic properties of encrypted data. The ciphering unit has not to be inserted between the cache memory and the processor otherwise the latency induced will be too important.

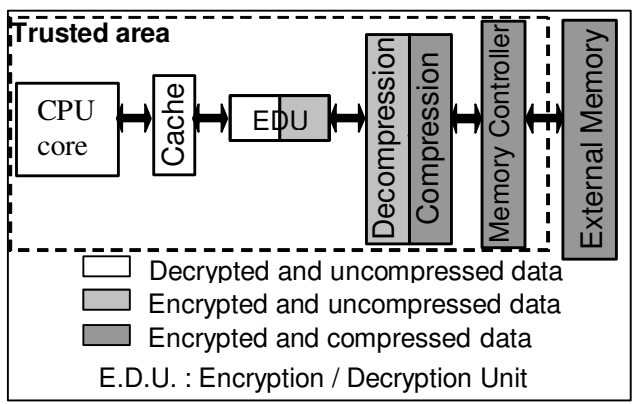

Figure 8: Compression and encryption

Compression can improve the performance of the encryption unit by decreasing the data size to cipher and to decipher. In addition, compression can raise hopes for a gain of memory capacity, and also performance benefit due to lowered bus usage. IBM proposes a tool for code compression: CodePack [16]. The performance impact is claimed to be about +/- 10% (depends on the type of memory used) and an increase of memory density of 35%.

Moreover, compression increases the message entropy and thus improves the efficiency of an encryption algorithm on the same message. Another benefit is that compression adds a layer of security.

## 5. Conclusion and perspectives

Designing a system offering a sufficient level of security, and as a result ensuring confidentiality is today feasible, thanks to a ciphering unit. Avoiding significant performance losses is the challenge. The first proposed scheme (insert the ciphering engine between the cache memory and CPU) appears to be difficult to set up, due to the sensitive character of the CPU-cache memory communication in term of timing. Moreover, doubling the integrated memory size seems to be unaffordable.

Consequently, one investigation proposed is to add a compression step preceding the ciphering step and eventually merge both on the same core. This might lower the performance loss and increase the security as well.

In future exploration, it might also be relevant to take into account the problem of integrity, to thwart attacks based on the modification of the fetched instructions.